\documentclass[journal]{IEEEtran}
\usepackage{amsmath, amssymb}

\usepackage{color}
\newcommand{\change}[1]{\textcolor{red}{#1}}

\newcounter{dum}

\newenvironment{mylist}{
\setcounter{dum}{1}
\begin{list}
{\arabic{dum}.}
{ \usecounter{dum}
  \setlength{\itemindent}{-3ex}
  \setlength{\labelsep}{0.5ex}
  \setlength{\labelwidth}{-1ex}
  \setlength{\leftmargin}{3ex}
  \setlength{\rightmargin}{0in}
  \setlength{\itemsep}{-0.2ex}}  }
{\end{list}}

	{\end{tabular}\end{center}}

\newcommand{\be}{\begin{equation}}
\newcommand{\ee}{\end{equation}}

\newcommand{\bea}{\begin{eqnarray}}
\newcommand{\eea}{\end{eqnarray}}

\newcommand{\bfg}{\begin{figure}[htbp]}
\newcommand{\efg}{\end{figure}}

\newcommand{\bd}{\begin{description}}
\newcommand{\ed}{\end{description}}

\newcommand{\ben}{\begin{itemize}%
	\setlength{\parsep}{0pt}\setlength{\itemsep}{-1pt}}
\newcommand{\een}{\end{itemize}}

\def\benum{\begin{enumerate}\setlength{\itemsep}{-0.6ex}\setlength{\parsep}{0pt}}
\def\eenum{\end{enumerate}}

\def\btab{\begin{tabbing}}
\def\etab{\end{tabbing}}

\def\mn{\medskip\noindent}
\def\mnb#1{\medskip\noindent{\bf #1}}

\def\tr{{\rm tr}}

\newcommand{\eq}[1]{Eq.~(\ref{eq:#1})}

\newcommand{\lem}[1]{Lemma~\ref{lem:#1}}

\def\lpm{ \left(\rule{0pt}{2.1ex}\right. }
\def\rpm{ \left.\rule{0pt}{2.1ex}\right) }

\def\lps{ \left(\rule{0pt}{1.8ex}\right. }
\def\rps{ \left.\rule{0pt}{1.8ex}\right) }

\def\<{\langle}
\def\>{\rangle}

\def\ot{\otimes}

\def\ra{\rightarrow}

\def\N{{\cal N}}
\def\Ma{{\cal M}_{\rm a}}
\def\Mi{{\cal M}_{\rm i}}
\def\cS{{\cal S}}
\def\e{\epsilon}
\def\eps{\epsilon}
\def\bbC{\mathbb{C}}

\newtheorem{theorem}{Theorem}
\newtheorem{lemma}[theorem]{Lemma}

\newcommand{\smfrac}[2]{\mbox{$\frac{#1}{#2}$}}

\newcommand{\bra}[1]{\mbox{$\left\langle #1 \right|$}}
\newcommand{\ket}[1]{\mbox{$\left| #1 \right\rangle$}}
\newcommand{\braket}[2]{\mbox{$\langle #1 | #2 \rangle$}}

\newcommand{\proj}[1]{\mbox{$| #1 \>\< #1 |$}}

\def\CcE{C_{{\rm cap}}^{{\rm EPR}}} 
\def\CcG{C_{{\rm cap}}^{{\rm ent}}} 

\def\CseE{C_{{\rm sim},\eps}^{{\rm EPR}}} 
\def\CseG{C_{{\rm sim},\eps}^{{\rm ent}}} 
\def\CceE{C_{{\rm cap},\eps}^{{\rm EPR}}} 
\def\CceG{C_{{\rm cap},\eps}^{{\rm ent}}} 

\def\CcfG{C_{{\rm cap},\rightarrow}^{{\rm ent}}} 
\def\CcbG{C_{{\rm cap},\leftarrow}^{{\rm ent}}}

\begin{document}

\title{A communication-efficient nonlocal measurement with
application to communication complexity and bipartite gate
capacities}

\author{Aram W. Harrow and Debbie W. Leung
\thanks{Aram W. Harrow is with the
Department of Mathematics, University of Bristol, Bristol, BS8 1TW, U.K.
and was funded by
the Army Research Office under grant
W9111NF-05-1-0294, the E.C. under grants
ASTQIT (FP6-022194) and QAP (IST-2005-15848), and the U.K. Engineering
and Physical Science Research Council through ``QIP IRC.''
{\em email:} {\tt a.harrow@bris.ac.uk}}
\thanks{Debbie W. Leung is with the
   Department of Combinatorics and Optimization, and 
   Institute for Quantum Computing, University of Waterloo, 
   Waterloo, Ontario, N2L 3G1, Canada, and was 
funded by the CRC, ORF, NSERC, CIFAR, MITACS, 
and QuantumWorks.  
{\em email:} {\tt  wcleung@iqc.ca}
}}

\date{\today}
\maketitle

\begin{abstract}
  Two dual questions in quantum information theory are to determine
  the communication cost of simulating a bipartite unitary gate, and
  to determine their communication capacities.  We present a bipartite
  unitary gate with two surprising properties:
  1) simulating it with the assistance of unlimited EPR pairs requires
  far more communication than with a better choice of entangled state,
  and 2) its communication capacity is far lower than its capacity to
  create entanglement.  This suggests that 1) unlimited EPR pairs are
  not the most general model of entanglement assistance for
  two-party communication tasks, and 2) the entangling and communicating
  abilities of a unitary interaction can vary nearly independently.
  The technical contribution behind these results is a
  communication-efficient protocol for measuring whether an unknown
  shared state lies in a specified rank-one subspace or its orthogonal
  complement.
\end{abstract}



\mnb{Introduction.}  Many basic questions in quantum information
theory can be phrased as determining the rates at which standard
communication resources (EPR pairs, noiseless qubit channels, etc.)
can be converted to and from more specialized resources (such as an
available noisy channel, or computation of functions of interest with
distributed inputs).  Typically local operations are allowed for free;
sometimes entanglement is as well.  For example, channel capacities
are the maximum rates at which noisy channels can be turned into
noiseless ones, while the quantum communication complexity of a
function $f$ is related to the minimum rate at which noiseless quantum
communication is turned into evaluations of $f$.

In quantum mechanics, the most general interaction between two
systems, given sufficient isolation from the environment, is a
bipartite unitary quantum gate $U$.  We will think of the systems ($A$
and $B$) as each comprising $n$ qubits, and as being held by two
parties, Alice and Bob. 
 
A fundamental goal of quantum information processing is to simulate
interactions (i.e.\ unitaries) using as few resources as possible.
This paper investigates these simulation costs when different types
of entanglement are given for free.  We will define $\CseG(U)$ to be
the number of bits of classical communication necessary to simulate
$U$ up to error $\eps$ if Alice and Bob are allowed to start with an
entangled state of their choice.
(Given free entanglement, the quantum and classical communication
costs differ by a factor of exactly $2$, due to teleportation
\cite{Bennett93} and super-dense coding \cite{Bennett92}.)
The canonical form of entanglement is the EPR pair, since it can be
converted to many copies of any other state using an asymptotically
vanishing amount of communication per copy \cite{BBPS}.  
Accordingly, we also let  $\CseE(U)$ denote the classical communication cost of
simulating $U$ up to error $\eps$ given unlimited EPR pairs.

Also of interest is the effectiveness of unitaries at sending
classical messages or generating entanglement.
The ultimate limit to which this can be done is given by the rate
achievable with an asymptotically large number of uses and vanishing
error (previously defined in \cite{BHLS02}).
Note that these unitaries can communicate in either direction, or both 
simultaneously.  We are primarily interested in the combined rate in 
both directions (as with simulation costs). 
Let $\CceG(U)$ and $\CceE(U)$ denote the largest number of bits that
$U$ can transmit {\em in a single use} up to error $\eps$, when allowed
arbitrary entanglement or free EPR pairs, respectively.
The corresponding asymptotic capacities are denoted $\CcG(U)$ and
$\CcE(U)$.
(Previous works \cite{BHLS02,HS05} used the notation
$C_+^E(U)$ for the latter scenario.) 
Likewise, let $E_{\text{cap}}(U)$ denote the asymptotic entanglement 
capacity.  
Naturally, simulation costs are upper bounds to communication
capacities.

We might reasonably expect that these capacities reflect the interaction
strength of the unitaries, and thus if one capacity is large, the
others should be as well.  For example, a gate that communicates well
in the forward direction ought to also do so in the backward direction, and a
highly entangling gate should also disentangle or communicate a lot.
This is indeed the case for some
well-studied unitaries (e.g., {\sc cnot}, {\sc swap}, and unitaries
close to the identity).
Additionally, it has been proven that if one of these capacities is
positive, the 
others are as well \cite{BHLS02},
and that communication capacities are generally lower bounds of the
entanglement capacity ($\CcG(U) = \CcE(U) \leq E_{\text{cap}}(U)+
E_{\text{cap}}(U^\dagger)$) \cite{BHLS02,BS03}.
However, beyond the above proven bounds, little support was found for
the intuition.
More recently, Ref.\ \cite{HS05} finds gates exhibiting arbitrarily
large differences between entanglement and disentanglement capacities,
(see also \cite{LSW05}), and between forward and backward communication
capacities.
In this paper, we demonstrate the remaining separation: an arbitrarily
large difference between entanglement capacity and communication
capacity.
Together with the results of \cite{HS05}, this indicates that most
unitary gate capacities of interest can vary nearly independently.

\mnb{The gate $U$.}
For our gate $U$, $A$ and $B$ each have $d{+}1$
dimensions (or equivalently, $n = \log (d{+}1)$ qubits) and a basis
given by $\{|0\>,\cdots,|d\>\}$.  Let $|\Phi\> = \smfrac{1}{\sqrt{d}}
\lps |11\> + \cdots + |dd\> \rps$ and $P=|00\>\<00| + |\Phi\>\<\Phi|$.
  Define 
$$U = |00\>\<\Phi| + |\Phi\>\<00| + I-P.$$
  In other
words, $U$ swaps $|00\>$ with $|\Phi\>$ and leaves the rest of the
space (i.e. the support of $I-P$)
unchanged.  Note that $U = U^\dagger$.  

We consider this gate $U$ because it can certainly create or remove
$\log d \approx n$ ebits but it leaves most of the space unchanged.
This latter property will allow us to simulate $U$ with little
communication, implying upper bounds on its communication capacity.  

\mnb{The simulation protocol $W$.$\;$} 
Define $|\phi_-\> = \smfrac{1}{\sqrt{2}} \lps |\Phi\> - |00\>\rps$.
Note that $U$ has only $1$ nontrivial eigenvalue, $-1$, and the
corresponding eigenvector is $|\phi_-\>$.
Let ${\cal M}_{\rm i}$ be the ideal coherent measurement that maps
$|\phi_-\>|0\> \ra |\phi_-\>|0\>$ and $|\phi\>|0\> \ra |\phi\>|1\>$
if $\<\phi|\phi_-\>=0$.  ${\cal M}_{\rm i}$ is a $2$-outcome
measurement with POVM elements
$M_0=|\phi_-\>\<\phi_-|, M_1=I-\proj{\phi_-}$.
The protocol $W$ simulates $U$ by using a nonlocal state
identification procedure $\Ma$ (described below) that will make use of
$|\phi_-\>^{\ot m-1}$ to approximate $\Mi$.  $W$ has 5 steps:
\begin{mylist}

\item Adjoin ancillas $|\phi_-\>^{\otimes m-1}$.

\item 
Apply ${\cal M}_{\rm a}$.  
Store the outcome $0/1$  in a qubit $C$ in Bob's possession (WLOG).
We will prove later that ${\cal M}_{\rm a}$ differs from ${\cal
  M}_{\rm i}$ in the diamond norm \cite{kitaev} by no more than
$O(m^{-1/2})$ using the catalyst $|\phi_-\>^{\otimes m-1}$ and $\log
(m)$ qubits of communication in each direction.

\item
Apply the gate ${\rm Diag}(-1,1)$ to $C$, so that
$\ket{0}$ is mapped to $-\ket{0}$ and $\ket{1}$ mapped to $\ket{1}$.

\item
Reverse ${\cal M}_{\rm a}$ in step 1, so as to coherently
erase the outcome in $C$.  This step also requires $\log (m)$
qubits of communication in each direction.

\item 
Discard the ancillas and system $C$.
\end{mylist}

\mnb{Procedure for nonlocal state identification ${\cal M}_{\rm a}$.}
We start with an informal description of the task, ignoring locality
constraints.  Suppose we want to know whether or not an unknown
incoming state $|\beta\>$ is equal to some other state $|\alpha\>$,
and we have possession of $m{-}1$ copies of $|\alpha\>$.  One
(approximate) method
is to project $|\alpha\>^{\otimes m{-}1} |\beta\>$ onto the symmetric
subspace of $(\bbC^d)^{\ot m}$ (defined as the span of all vectors of
the form $\ket{\psi}^{\ot m}$ for $\ket{\psi}\in\bbC^d$; see Ref.~\cite{BBDEJM96} for more background).  This
defines a two-outcome measurement with measurement operators
$\Pi_{\text{sym}}:=\smfrac{1}{m!}  \sum_{\pi\in \cS_{m}} \!\! \pi$,
and $I-\Pi_{\text{sym}}$.  (Here $\cS_{m}$ is the group of operators
that permute the $m$ registers.)
The outcome corresponding to $\Pi_{\text{sym}}$ occurs with
probability 
$\bra{\alpha}^{\otimes m{-}1}\bra{\beta}
\; \smfrac{1}{m!} \sum_{\pi\in\cS_m} \!\! \pi \, 
\ket{\alpha}^{\ot m{-}1}\ket{\beta}$. 
%
A fraction $\smfrac{1}{m}$ of the permutations fix the $m^{\text{th}}$
register. 
For each such $\pi$, $\bra{\alpha}^{\otimes m-1}\bra{\beta} \pi
\ket{\alpha}^{\ot m-1}\ket{\beta}= 1$.
The remaining $1{-}\smfrac{1}{m}$ fraction of the permutations swaps
the $m^{\text{th}}$ register with one of the others.  In this case
$\bra{\alpha}^{\otimes m-1}\bra{\beta} \pi \ket{\alpha}^{\ot
m-1}\ket{\beta}= |\braket{\alpha}{\beta}|^2$.  Thus the 
probability of obtaining $\Pi_{\text{sym}}$ is $\smfrac{1}{m} +
(1{-}\smfrac{1}{m})|\braket{\alpha}{\beta}|^2 = |\<\alpha|\beta\>|^2 +
\smfrac{1}{m} (1 {-} |\<\alpha|\beta\>|^2)$, and the procedure
simulates the measurement 
with operators $\{\proj{\alpha},I-\proj{\alpha}\}$ up to error
at most $1/m$.

Observe that instead of $\pi$ ranging over all $m!$ permutations, it
would suffice to take only the $m$ cyclic permutations.  For the
multi-partite setting, this will allow us to save dramatically on
communication.
We now describe the bipartite protocol and {a careful bound on the
accuracy is derived in the appendix}.

Let $|s\> = \smfrac{1}{\sqrt{m}} \sum_{j=0}^{m-1} |j\>$ and 
$S$ be a register prepared in the state $|s\>$.  Let $Y$ act on $S
\otimes (\bbC^{{d+1}})^{\ot m}$ by mapping $|j\> |\psi_1\> |\psi_2\> \cdots
|\psi_m\>$ to $|j\> |\psi_{1{-}j}\> |\psi_{2{-}j}\> \cdots
|\psi_{m{-}j}\>$, with arithmetic done mod $m$.  That is, $S$ controls a
cyclic permutation of the $m$ registers, taking the first register to the
$(j+1)^{\text{st}}$ one if the state of $S$ is $|j\>$.

With a slight abuse of notation, let ${\cal M}_{\rm i}$ and ${\cal
M}_{\rm a}$ be the ideal and approximate coherent state identification
protocols for some bipartite state $|\alpha\>$, with the answer
residing with Bob.  The state to be measured lives in systems $AB$.
Alice and Bob already share $|\alpha\>^{\otimes m-1}$ in $A_2 B_2
\otimes \cdots \otimes A_m B_m$.  ${\cal M}_{\rm a}$ is given by:
\begin{mylist}
\item Alice prepares a register $S$ in the state $|s\>$. 
\item Alice applies $Y$ on $S \otimes A \otimes A_2 \cdots A_m$
(i.e. she applies the $S$-controlled cyclic permutation on her halves
of the $m$ bipartite systems).
\item Alice sends $S$ to Bob using $\log(m)$ qubits of forward
communication. 
\item Bob performs $Y$ on $S \otimes B \otimes B_2 \cdots B_m$ thereby
completing the $S$-controlled cyclic permutation on the $m$ bipartite
systems.
\item Bob coherently measures $S$ with POVM $\{|s\>\<s|, I-|s\>\<s|\}$. 
The final outcome is written to a register $C$ in Bob's possession.  
\item Bob performs $Y^\dagger$ on $S \otimes B \otimes B_2 \cdots B_m$.
\item Bob sends $S$ to Alice using $\log(m)$ qubits of backward
communication. 
\item Alice applies $Y^\dagger$ on $S \otimes A \otimes A_2 \cdots A_m$.
\end{mylist}

{We quantify the accuracy of the simulation using 
the diamond-norm, which, for a superoperator ${\cal S}$, is
defined as $\|{\cal S}\|_\diamond {:}{=} \max_{\psi \geq 0, \tr \psi =
1} \| ({\cal I}\ot {\cal S})(\psi) \|_1$. 
In particular, we prove (in the appendix) that:
\begin{theorem} \label{thm:dianorm}
~$\| {\cal M}_{\rm a} - {\cal M}_{\rm i}
\|_\diamond \leq \smfrac{{2}\sqrt{2}}{\sqrt{m}}$.  
\end{theorem}
}

{Now, in} the protocol $W$ that simulates $U$, if we
replace the two uses of ${\cal M}_{\rm a}$ by ${\cal M}_{\rm i}$, we
obtain an exact implementaion of $U$.  By the triangle inequality, $\|
U - W\|_\diamond \leq 2 \, \| {\cal M}_{\rm a} - {\cal M}_{\rm i}
\|_\diamond \leq \smfrac{{4}\sqrt{2}}{\sqrt{m}}$.  For $W$ to simulate
$U$ with accuracy $\e$, it suffices to take $m = \smfrac{{32}}{\e^2}$.  The
simulation consumes $2 \log m$ qubits of communication in each direction.  
Thus we have the following.

\begin{theorem} \label{thm:simcost}
~$\CseG(U) \leq {40} + 16 \log \smfrac{1}{\e}$.  \\[-2ex]
\end{theorem}


Note that $U$ is implicitly parameterized by the system size $n$, yet
the simulation cost is independent of it.  Next we prove two results
based on the simulation protocols and Theorem \ref{thm:simcost}.


\mnb{Consequence 1: {Simulation with EPR pairs can be suboptimal}}

\begin{theorem} $\forall \epsilon>0$, 
\label{thm:EPR-sim-LB}
\be \CseE(U) \geq 
2\log(d)-1+\log((1-2\delta)(1-\delta)^2) \,,
\nonumber
\ee
where $\delta := \sqrt[8]{2\eps}$. 
\end{theorem}

\mn{\it Proof.}  Let $A'B'$ denote auxiliary systems held by Alice and
Bob.  Consider the transformation of an arbitrary state
$\ket{\varphi_1}_{AA'BB'}$ to $\ket{\varphi_2}_{AA'BB'} = U_{AB}
\otimes I_{A'B'} \ket{\varphi_1}_{AA'BB'}$.  The communication cost to
perform this transformation with high fidelity is a lower bound on the
communication cost to approximately simulate the gate $U$, assuming
that EPR pairs are free in both scenarios.  
Let 
$\rho_{1,2} = \tr_{BB'} |\varphi_{1,2}\>\<\varphi_{1,2}|$. 

Corollary 10 of Ref~\cite{HW02} states that if $\ket{\varphi_1}_{AA'BB'}$
can be transformed to $\ket{\varphi_2}_{AA'BB'}$ with fidelity 
$(1{-}\kappa)^{1/2}$ 
by exchanging a total of $C$ classical bits and consuming EPR pairs, then, 
$C \geq \Delta_\delta(\rho_2) - \Delta_0(\rho_1) + 2 \log(1{-}\delta)$ 
where $\delta = (4 \kappa)^{1/8}$, and 
$\Delta_\delta(\rho) = \log \min_J [|J| \max(J)]$, where 
$J$ is any subset of eigenvalues of $\rho$ whose entries sum to at
least $1{-}\delta$, $|J|$ is the size of the set, and $\max(J)$ is the
maximum element of $J$.
This statement is based on a definition of fidelity as
$F(\sigma,\omega) = \tr\sqrt{\sigma^{1/2} \omega \sigma^{1/2}}$ which
is the square-root of that defined in Ref~\cite{HW02}.
When one of the states is pure, the fidelity satisfies the relation
$1-F(\sigma,\omega)^2 \leq \smfrac{1}{2} \|\sigma-\omega \|_1$.  
When Alice and Bob apply to $|\varphi_{1}\>$ an approximate simulation
of $U$ with accuracy $\epsilon$ in the diamond norm, the output state
is $\epsilon$ close to $\ket{\varphi_2}$ in $1$-norm.  So, this
achieves an approximate transformation of $|\varphi_{1}\>$ to
$|\varphi_{2}\>$ with fidelity at least
$(1-\smfrac{\epsilon}{2})^{1/2}$.  Thus, the corollary applies with
$\kappa = \smfrac{\epsilon}{2}$ and $\delta = (4 \kappa)^{1/8} = (2
\epsilon)^{1/8}$.

Recall that $\ket{\Phi} = \smfrac{1}{\sqrt{d}} (|11\>+\cdots+|dd\>)$.
We take $\ket{\varphi_1} = \frac{1}{\sqrt 2}(\ket{\Phi}_{AB}\ot
\ket{00}_{A'B'} + \ket{00}_{AB}\ot \ket{\Phi}_{A'B'})$, thus
$\ket{\varphi_2} = \frac{1}{\sqrt 2}(\ket{00}_{AB}\ot \ket{00}_{A'B'}
+ \ket{\Phi}_{AB}\ot \ket{\Phi}_{A'B'})$.
$\ket{\varphi_1}$ is a maximally entangled state of Schmidt rank $2d$.
Thus, $\Delta_0(\rho_1) = 0$.  
The state $\rho_2$ has a nondegenerate eigenvalue $1/2$, and a
degenerate one $\smfrac{1}{2d^2}$ with multiplicity $d^2$.  The
optimal $J$ has $|J| = 1+d^2 - \lfloor 2 \delta d^2 \rfloor$ and
$\max(J) = 1/2$.  Therefore, $\Delta_\delta(\rho_2) \geq \log
[(1+(1{-}2 \delta) d^2)/2]$ $\geq \log (1{-}2 \delta) + 2 \log d -1$.
Substituting $\Delta_\delta(\rho_2)$ and $\Delta_0(\rho_1)$ into 
the corollary gives the stated lower bound on the communication cost.
$\hfill \square$

Comparing Theorems \ref{thm:simcost} and \ref{thm:EPR-sim-LB}, for
constant $\epsilon \ll 1$, the simulation cost is $\approx 2 \log d$ given
unlimited EPR pairs and $\approx 16 \log \smfrac{1}{\epsilon}$ when
$O(\smfrac{1}{\e^2})$ copies of $\ket{\phi_-}$ are available.

Note that {\em any} $n\times n$-qubit unitary can be trivially
simulated with EPR pairs and $4n$ bits of communication by teleporting
Alice's input to Bob, having him apply the unitary and then
teleporting her system back.  Thus, Theorem \ref{thm:EPR-sim-LB}
implies that even given unlimited EPR pairs and allowing a small
error, simulating $U$ is at least half as costly as simulating a
completely general unitary on $n\times n$ qubits.


\mnb{Consequence 2: Some gates can entangle exponentially more than they
can communicate.} \\[1ex]
Since $U\ket{00}=\ket\Phi$, we can bound
$E_{\rm cap}(U) \geq \log (2^n-1) \approx n$.   
On the other hand, we have:
\begin{theorem}\label{thm:xoxo}
For all $n$, $\CcG(U)\leq
16\log n+100$.
\end{theorem}

When communicating using a gate in both directions simultaneously,
there is generally a tradeoff between the forward and backward
communication rates.  The one-way capacity in each direction is an
extreme point of that tradeoff.  We denote these capacities by
$\CcfG(U)$ and $\CcbG(U)$.  Theorem \ref{thm:xoxo} can be proved by showing
$\CcfG(U) \leq 8\log n+50$, since the symmetry of $U$ means that the
same bound applies to $\CcbG(U)$, and finally we can bound $\CcG(U)
\leq \CcfG(U) + \CcbG(U) \leq 16 \log n + 100$.

\mn{\it Proof of }$\CcfG(U) \leq 8 \log n + 50$.  

\vspace*{1ex}

The nonlocal state identification protocol ${\cal M}_{\rm a}$ uses
shared entangled states between Alice and Bob and $\log m$ qubits of
communication in each direction, and the protocol $W$ that simulates
$U$ uses ${\cal M}_{\rm a}$ twice, $W$ uses $2 \log m$ qubits of
forward communication.  But back communication and shared entanglement
cannot increase the classical capacity of a noiseless forward quantum
channel beyond the superdense-coding bound \cite{CDNT97}, thus
 \be \hspace*{-2ex} \CcfG(W)\leq 4\log m \,. \ee
It remains to show that $\CcfG(W) \approx \CcfG(U)$ if
$\|W-U\|_\diamond$ is small.  To make this quantitative, 
we prove the following {\it continuity bound} in the appendix.
\begin{lemma}\label{lem:continuity}
If $\N_1$, $\N_2$ are bidirectional channels with outputs in
$\mathbb{C}^{d{+}1} \ot \mathbb{C}^{d{+}1}$ such that $\|\N_1 -
\N_2\|_\diamond \leq \e$, then
\be | \CcfG(\N_1) - 
\CcfG(\N_2) | \leq 8 \e \log (d{+}1) + 4 H_2(\e)
\nonumber
\ee where $H_2$ is the
binary entropy function.
\end{lemma}

Our continuity bound means that the more accurate $\Ma$ is, the closer
the capacities of $U$ and $W$ are. On the other hand, making $\Ma$
more accurate requires more communication.  Thus we face a trade-off
between keeping the capacity of $W$ small and keeping the capacities
of $U$ and $W$ close to each other.  Optimizing will give us a bound
of $O(\log n)$ bits on the capacity of $U$.

\mn{\it Completing the proof of} $\CcfG(U) \leq 8 \log n+50$.

\vspace*{1ex}

Recall that the accuracy of the approximate nonlocal state
identification in terms of the communication cost is $\eta =
\smfrac{\change{2} \sqrt{2}}{\sqrt{m}}$, and that $\| U {-} W\|_\diamond \leq 2
\eta = \e$.
According to \lem{continuity}, since $\log(d{+}1) = n$, the
difference in the capacities of $U$ and $W$ is suppressed if $m = n^c$
for $c>2$.
More precisely, 
\bea
\CcfG(U) & \hspace*{-1.4ex} \leq \hspace*{-1.4ex} 
         & \CcfG(W) + 16 \eta \log(d{+}1) + 4 H_2(2\eta) 
\nonumber
\\ & \hspace*{-1.4ex} \leq \hspace*{-1.4ex} 
         & 4 \log m + 16 \eta n + \min(8 \sqrt{2\eta},4)
\nonumber
\\ & \hspace*{-1.4ex} \leq \hspace*{-1.4ex} 
   & 4 c \log n + 32 \sqrt{2} n^{1- {c\over2} } 
    + \min(16 \cdot 2^{1 \over 4} n^{-{c\over4}},4)
\nonumber
\eea
where each term is bounded by the corresponding term in the subsequent
line (and $H_2(x) \leq \min(2 \sqrt{x},1)$).  For sufficiently large
$n$ and $c>2$, we have $\CcfG(U) \leq 4 c \log n$ and $\CcG(U)\leq 8c
\log n$.  For arbitrary $n$, choosing $c=2$ gives 
$ \CcfG(U) \leq  8 \log n + 32 \sqrt{2} + 4 \leq 8 \log n + 50.$
$\hfill$ $\square$

\mnb{Extensions.}

{
Our nonlocal state identification protocol generalizes
straightforwardly to more than two remote parties (say, $k$).  Two examples to consider are a cyclic network topology and a
star-shaped network.
In the cyclic topology, one party creates the state $|s\>$ as
defined before and $|s\>$ is then circulated among all parties.  In
the star-shaped network, the $k$ parties share $|s\> =
\smfrac{1}{\sqrt{m}}\sum_{j{=}0}^{m-1} |j\>^{\otimes k}$, each sends
his share to the party designated to have the answer, who returns
these shares to complete the protocol.
} 

Our {gate} simulation procedure allows us to simulate any bipartite gate
with $r$ non-trivial eigenvalues using $O(r\log(r/\eps))$ qubits of
communication.  This is accomplished by testing the state held by
Alice and Bob sequentially against each of the $r$ corresponding
eigenvectors.  Each individual test needs to have error $\eps/r$ so
that the total error can be bounded by $\eps$.  This simulation method
is useful for $r \ll \log(d)$ (since a gate can be trivially simulated
using $\log d$ qubits of communication in each direction).
{It will be interesting to find better simulation protocols
for large $\log(d) \ll r \ll d$.}


Regarding unitary gate capacities, we have shown that $\CcG(U)$ can
scale like the logarithm of $E_{\text{cap}}(U)$.  However, it is
unknown how much further this result could be improved.  For our
example, it is possible that $\CcG(U)$ can be upper-bounded by a
constant even as $n\ra \infty$.  Moreover, it is possible that even
stronger separations are possible.  Bound 1 of \cite{BHLS02} implies
that $\CcG(U)>0$ whenever $E_{\text{cap}}(U)>0$, but even for fixed
dimension no nonzero lower bound on $\CcG(U)$ is known.  The
difficulty is that the proof in \cite{BHLS02} relates $\CcG(U)$ to 
the amount of entanglement which one use of $U$ can create from
unentangled inputs.  This quantity can be arbitrarily smaller than
$E_{\text{cap}}(U)$ even for fixed dimensions.

\mnb{Acknowledgements.}
We are grateful to Charles Bennett, whose hope for a simple theory
concerning interconversions between nonlocal resources has inspired
many of our investigations in this subject.


\onecolumn\newpage
\appendix
\section{Proofs}
\mnb{A. Proving that 
$\|{\cal M}_{\rm a}-{\cal M}_{\rm i}\|_\diamond \leq \smfrac{2\sqrt{2}}{\sqrt{m}}$.} 

We use the notations that are defined in the main text.  

To upper bound $\| {\cal M}_{\rm a} - {\cal M}_{\rm i} \|_\diamond$,
the most general initial state can be expressed as 
$$|\phi\> = \sqrt{p} \, |a\>_R |\alpha\>_{AB} + \sum_i \sqrt{p_i} \,
|a_i\>_R |\alpha_i\>_{AB},$$
where $R$ is a reference system that may be entangled with the
incoming systems $AB$, the states $|\alpha_i\>_{AB}$ and
$|\alpha\>_{AB}$ form a basis on $AB$, and $|a\>_R, |a_i\>_R$ are unit
vectors that are not necessarily orthogonal to one another, $p, p_i
\geq 0$ and $p + \sum_i p_i = 1$.

We now analyze how each step in ${\cal M}_{\rm a}$ evolves $|\phi\>$.
We will include all the auxiliary systems in the analysis, and each of
these steps is coherent.  Thus, we are analyzing the isometric
extensions of ${\cal M}_{\rm a}$ and ${\cal M}_{\rm i}$ 
%
%
and it suffices to keep track of the pure state over all the relevant
systems.
Our goal is to express the final state $|{\rm fin}\>$ as a sum of the
``correct state'' $|{\rm cor}\>$ (obtained by coherently applying ${\cal
  M}_{\rm i}$ to $|\phi\>$) and an error term $|{\rm err}\>$.

The state after attaching the ancillas (step 1) is: 
$$ \sqrt{p} \, |a\> |\alpha\>^{\otimes m} |s\> + \sum_i \sqrt{p_i} \,
|a_i\> |\alpha_i\> |\alpha\>^{\otimes m{-}1} |s\> \,.$$
After Alice applies $Y$, communicates $S$ to Bob, and Bob applies $Y$ (steps 2-4), 
the state becomes: 
$$ \sqrt{p} \, |a\> |\alpha\>^{\otimes m} |s\> + 
 \sum_i \sqrt{p_i} \, |a_i\> \smfrac{1}{\sqrt{m}} \sum_{j=0}^{m-1}
|\alpha\>^{\otimes j}|\alpha_i\> |\alpha\>^{\otimes m{-}1{-}j} |j\>\,.$$
In step 5, Bob attaches $|0\>_C$ and makes the coherent measurement on $S$,
taking $|s\>|0\>_C \ra |s\>|0\>_C$ and $|s_\perp\>|0\>_C \ra
|s_\perp\>|1\>_C$ for all $\<s_\perp|s\>=0$.  
To write down the resulting state, we should rewrite each $|j\>$ in
the Fourier basis which includes $|s\>$.  But to obtain just a bound,
we can simply express $|j\> = \smfrac{1}{\sqrt{m}} |s\> +
\smfrac{\sqrt{m{-}1}}{\sqrt{m}} |s_j\>$ where $\<s_j|s\>=0$ for each $j$.
The measurement on $S$ thus results in the state
$$ \sqrt{p} \, |a\> |\alpha\>^{\otimes m} |s\> |0\> + 
\sum_i \sqrt{p_i} \, |a_i\> \smfrac{1}{\sqrt{m}} \sum_{j=0}^{m-1}
|\alpha\>^{\otimes j}|\alpha_i\> |\alpha\>^{\otimes m{-}1{-}j} 
\lpm \smfrac{1}{\sqrt{m}} 
|s\>|0\> + \smfrac{\sqrt{m{-}1}}{\sqrt{m}} |s_j\>|1\> \rpm .$$
Here, the second occurrence of the $|s\>|0\>$ term (the one in the
parenthesis) represents an erroneous measurement outcome.  We add and
subtract $\smfrac{1}{\sqrt{m}} |s\>|1\>$ in the parenthesis: 
$$ \sqrt{p} \, |a\> |\alpha\>^{\otimes m} |s\> |0\> + 
\sum_i \sqrt{p_i} \, |a_i\>  \smfrac{1}{\sqrt{m}} \sum_{j=0}^{m-1} 
|\alpha\>^{\otimes j}|\alpha_i\> |\alpha\>^{\otimes m{-}1{-}j} 
\lpm \smfrac{1}{\sqrt{m}} 
|s\>(|0\>{-}|1\>) + |j\>|1\> \rpm .$$
Rearranging, we get:  
$$ \sqrt{p} \, |a\> |\alpha\>^{\otimes m} |s\> |0\> 
+ \sum_i \sqrt{p_i} \, |a_i\> \smfrac{1}{\sqrt{m}} \sum_{j=0}^{m-1}
|\alpha\>^{\otimes j}|\alpha_i\> |\alpha\>^{\otimes m{-}1{-}j} 
|j\>|1\>$$
\vspace*{-2ex}
$$ + \sum_i \sqrt{p_i} \, |a_i\> \smfrac{1}{\sqrt{m}} \sum_{j=0}^{m-1} 
|\alpha\>^{\otimes j}|\alpha_i\> |\alpha\>^{\otimes m{-}1{-}j} 
\smfrac{\sqrt{2}}{\sqrt{m}} |s\>|-\> 
$$
where the first line is what an ideal measurement will produce (with
unit norm), and the second line represents an error term (and it is
{\em not} orthogonal to the ideal term, since the sum is also
normalized).
Now, Bob applies $Y^\dagger$ and sends $S$ back to Alice, who then
applies $Y^\dagger$ (steps 6-8), resulting in the final state $|{\rm fin}\> = 
|{\rm cor}\> + |{\rm err}\> $ where 
$$ |{\rm cor}\> = 
\sqrt{p} \, |a\> |\alpha\>^{\otimes m} |s\> |0\> + 
 \sum_i \sqrt{p_i} \, |a_i\> |\alpha_i\> |\alpha\>^{\otimes m{-}1} |s\> |1\>
$$
$$ |{\rm err}\> =  \smfrac{\sqrt{2}}{m^{3/2}} \, 
  \sum_i \sqrt{p_i} \, |a_i\> 
  \sum_{j,j'=0}^{m-1}  
 |\alpha\>^{\otimes  j{-}j'} 
 |\alpha_i\>
 |\alpha\>^{\otimes  m{-}1-(j{-}j')} |j'\> |-\>
$$
and as a reminder, the systems from left to right are $R$, $AB$, 
$A_2B_2$, $\cdots$, $A_mB_m$, $S$, and $C$.  

Next, we explicitly calculate $|\<{\rm cor}|{\rm err}\>|$.  
Replacing the dummy index $i$ by $i'$ in $|{\rm cor}\>$, and using the 
fact $|\alpha\>$ and $|\alpha_i\>$'s form a basis, only the $i=i'$ and 
$j=j'$ terms contribute to the inner product, which is 
$\smfrac{\sqrt{2}}{m^{3/2}} \sum_i p_i \sum_{j=0}^{m-1} \<s|j\> \<1|-\> 
= \smfrac{1}{m} \sum_i p_i = \smfrac{1{-}p}{m}$.
This implies 
$|\<{\rm cor}|{\rm fin}\>| \geq 1 -
{\smfrac{1{-}p}{m} \geq 1 -
\smfrac{1}{m}}$.
We are now ready to apply the well known relation 
$$\smfrac{1}{2} 
\| \; |a\>\<a| - |b\>\<b| \; \|_1 = \sqrt{1-|\<a|b\>|^2} \leq 
\sqrt{2 \, (1{-}|\<a|b\>|)} $$
to bound $\|{\cal M}_{\rm a} - {\cal M}_{\rm i} \|_\diamond$ which is 
equal to 
\bea
\nonumber
& = & \sup_{|\phi\>} \| ({\cal I} \otimes {\cal M}_{\rm a}) (|\phi\>\<\phi|) 
-  ({\cal I} \otimes {\cal M}_{\rm i}) (|\phi\>\<\phi|) \|_1 
\\ 
& = & \sup_{|\phi\>} \| \; |{\rm cor}\>\<{\rm cor}| - |{\rm fin}\>\<{\rm fin}| 
\nonumber
\; \|_1 \leq \smfrac{{2}\sqrt{2}}{\sqrt{m}} \,.
\eea

\vspace*{2ex}

\noindent {\bf B. Proof of \lem{continuity}:} Our proof will closely parallel
that of Lemma 1 of 
\cite{HS05}, which is similar to the above but holds for the case when
$\N_1$ and $\N_2$ are isometries.  The main ingredient in both proofs
is a single-shot capacity formula for bidirectional channels, first
established for isometries in \cite{BHLS02}, but then extended to
arbitary bidirectional channels in \cite{CLL05}: 
%
%
\be \CcfG(W) =
\sup_{\rho^{XAA'BB'}} I(X;BB')_{W(\rho)} -
I(X;BB')_{\rho}.\label{eq:single-shot}\ee 
Here $A,B$ are the registers acted on by $W$, $A',B'$ are ancillas of
arbitrary dimension, $X$ is a classical register,
$I(X;Y)=H(X)+H(Y)-H(XY)$ is the quantum mutual information of the
state given by the subscript.  $H(R) = H(\sigma) = -\tr \sigma \log
\sigma$ is the von Neumann entropy for the reduced density matrix
$\sigma$ on the system $R$.
When one of the registers $X$ is classical, the state on $XY$
represents an ensemble of quantum states on $Y$ labeled by basis
states of $X$, and the quantum mutual information is the Holevo
information \cite{Holevo73d}. \eq{single-shot} can be interpreted to
mean that $\CcfG(W)$ equals the largest single-shot increase in
mutual information possible when applying $W$ to any ensemble of
bipartite states.
Due to \eq{single-shot}, 
\be \CcfG(U) - \CcfG(W) 
\leq I(X;BB')_{U(\rho)} - I(X;BB')_{W(\rho)}
\label{eq:diff}
\ee
where $\rho$ attains the supremum in the expression for $\CcfG(U)$
to some arbitrary precision.  (This precision parameter is independent
from all other parameters considered, and thus will be omitted for
simplicity.)

Thus the desired continuity bound is essentially a continuity result
for quantum mutual information.  The crucial challenge is the lack of
dimensional bounds on the systems $X$ and $B'$, so that Fannes  
inequality \cite{Fannes73} does not provide the needed continuity
result.  Instead, we use a generalization due to Fannes and Alicki 
\cite{FA} that applies to conditional entropy: 
$$|H(Y|Z)_\sigma - H(Y|Z)_{\sigma'}| \leq 4\eps\log d + 2H_2(\eps)
\,,$$ where $\eps=\|\sigma - \sigma'\|_1$ and $d=\dim Y$.  Remarkably,
this Fannes-Alicki inequality provides an upper bound that is
independent of the size of the conditioned system $Z$.

Returning to \eq{diff}, first note that if $\|W - U\|_\diamond \leq
\eps$, then $\|W(\rho)-U(\rho)\|_1\leq \eps$.
Next, we can expand $I(X;BB')$ as
$$I(X;BB') = H(B') + H(B|B') - H(B|B'X) - H(B'|X)\,.$$ 
We now bound the difference of each of the above terms when evaluated
on $W(\rho)$ and $U(\rho)$.
The $H(B')$ and $H(B'|X)$ terms are the same for both states since $W$
and $U$ act only on $A,B$.  Applying the Fannes-Alicki inequality to
the remaining two terms and using $\dim B = d+1$ establishes the
Lemma.


\end{document}